\input harvmac.tex
\let\includefigures=\iftrue
\newfam\black
\includefigures
\input epsf
\def\figin{\epsfcheck\figin}\def\figins{\epsfcheck\figins}
\def\epsfcheck{\ifx\epsfbox\UnDeFiNeD
\message{(NO epsf.tex, FIGURES WILL BE IGNORED)}
\gdef\figin##1{\vskip2in}\gdef\figins##1{\hskip.5in}
\else\message{(FIGURES WILL BE INCLUDED)}%
\gdef\figin##1{##1}\gdef\figins##1{##1}\fi}
\def\DefWarn#1{}
\def\figinsert{\goodbreak\midinsert}
\def\ifig#1#2#3{\DefWarn#1\xdef#1{fig.~\the\figno}
\writedef{#1\leftbracket fig.\noexpand~\the\figno}%
\figinsert\figin{\centerline{#3}}\medskip\centerline{\vbox{\baselineskip12pt
\advance\hsize by -1truein\noindent\footnotefont{\bf Fig.~\the\figno:} #2}}
\bigskip\endinsert\global\advance\figno by1}
\else
\def\ifig#1#2#3{\xdef#1{fig.~\the\figno}
\writedef{#1\leftbracket fig.\noexpand~\the\figno}%
\global\advance\figno by1}
\fi
\def\co{\cal O}

\def\ka{K\"ahler}
\def\dm{\hbox{det}M}
\Title{\vbox{\baselineskip12pt\hbox{hep-th/9804059}
\hbox{RU-98-11, PUPT-1780,}
\hbox{ITEP-TH-52/97,}
\hbox{Landau-97-TMP-4}}}
{\vbox{
\centerline{Three Dimensional $N=2$ Gauge Theories and}
\vskip 10pt
\centerline{Degenerations of Calabi-Yau Four-Folds}}}
\vskip 10pt
\centerline{Duiliu-Emanuel Diaconescu$^\natural$
and Sergei Gukov$^\sharp$ \foot{On leave from
the Institute of Theoretical and Experimental
Physics and L.D. Landau Institute for Theoretical
Physics, Russia}}
\medskip
\centerline{\it ${}^\natural$ Department of Physics and Astronomy}
\centerline{\it Rutgers University }
\centerline{\it Piscataway, NJ 08855--0849}
\centerline{\tt duiliu@physics.rutgers.edu}
\medskip
\centerline{\it ${}^\sharp$ Joseph Henry Laboratories}
\centerline{\it Princeton University}
\centerline{Princeton, New Jersey 08544}
\centerline{\tt gukov@pupgg.princeton.edu}
\medskip
\bigskip
\noindent

Three dimensional $N=2$ gauge theories with arbitrary
gauge group and fundamental flavors are engineered from
degenerations of Calabi-Yau four-folds. We show how
Coulomb and Higgs branches emerge in the geometric picture.
The analysis of instanton generated superpotentials
unravels interesting aspects of the five-brane effective
action in M theory.

\bigskip\noindent{\vbox{
\hbox{key words: four-folds, gauge theories, degenerations}
\hbox{PACS numbers: 11.25.Mj, 11.30.Pb, 11.15.Tk}}}

\Date{April 1998}

\newsec{Introduction}

Recently, string theory has provided remarkable insights in
non-perturbative field theory dynamics.
It has turned out that sophisticated field theory phenomena
have a clear interpretation in terms of brane configurations
\nref\HW{A. Hanany, E. Witten, ``Type IIB Superstrings, BPS Monopoles,
And Three-Dimensional Gauge Dynamics'', Nucl.Phys. {\bf B492} (1997)
152, hep-th/9611230.}%
\nref\Wb{E. Witten, `` Solutions Of Four-Dimensional  Field Theories Via
M Theory'', Nucl.Phys. {\bf B500} (1997) 3, hep-th/9703166.}%
\nref\Kutasov{S. Elitzur, A. Giveon, D.Kutasov, ``Branes and $N=1$
duality in String Theory'', Phys. Lett. {\bf B400} (1997), 269,
hep-th/9702014; ``Brane Dynamics and $N=1$ Supersymmetric Gauge
theory'', Nucl.Phys. {\bf B505} (1997) 202, hep-th/9704104.}%
\refs{\HW,\Wb,\Kutasov}
or local degenerations of Calabi-Yau spaces
\nref\KV{S. Katz, A. Klemm, C. Vafa,
``Geometric Engineering of Quantum Field Theories'',
Nucl.Phys. {\bf B497} (1997) 173, hep-th/9609239.}%
\nref\KVa{S. Katz, C. Vafa, ``Geometric Engineering of N=1
Quantum Field Theories'', Nucl.Phys. {\bf B497} (1997) 196,
hep-th/9611090.}%
\nref\BJV{M. Bershadsky, A. Johansen, T. Pantev, V. Sadov, C. Vafa,
``F-theory, Geometric Engineering and N=1 Dualities'',
Nucl.Phys. {\bf B505} (1997) 153, hep-th/9612052.}%
\nref\OV{H. Ooguri, C. Vafa,
``Geometry of N=1 Dualities in Four Dimensions'', Nucl.Phys.
{\bf B500} (1997) 62, hep-th/9702180.}%
\refs{\KV,\KVa,\BJV,\OV}.
In certain situations, it is possible to construct a direct
correspondence between the two types of constructions
\ref\LV{N.C. Leung, C. Vafa, ``Branes and Toric Geometry'',
hep-th/9711013.}.

We consider $N=2$ supersymmetric gauge theories with fundamental
matter in three dimensions. As their $N=1$
four dimensional counterparts, these theories
exhibit very interesting phenomena such as
confinement, Seiberg duality, emergence of
various  phases and instanton effects
which require a deeper string theory understanding.
While brane and field theoretic aspects have been
already discussed in
\nref\BHOY{J. de Boer, K. Hori, Y. Oz and Z. Yin,
``Branes and Mirror Symmetry in N=2 Supersymmetric Gauge Theories
in Three Dimensions'',
Nucl.Phys. {\bf B502} (1997) 107, hep-th/9703100.}%
\nref\BHO{J. de Boer, K. Hori and Y. Oz,
``Dynamics of N=2 Supersymmetric Gauge Theories in Three
Dimensions'', Nucl.Phys. {\bf B500} (1997) 163, hep-th/9703100.}%
\nref\AISS{O. Aharony, A. Hanany, K. Intriligator, N. Seiberg,
M.J. Strassler,
"Aspects of N=2 Supersymmetric Gauge Theories in Three Dimensions'',
Nucl.Phys. {\bf B499} (1997) 67, hep-th/9703110.}%
\refs{\BHOY,\BHO,\AISS},
we focus on geometric constructions.

It is clear that three dimensional QCD with the desired supersymmetry
can be geometrically engineered from M theory compactification
on Calabi-Yau four-folds. Similar F theory compactifications on specific
compact Calabi-Yau spaces leading to four-dimensional theories
were investigated in
\nref\BJP{M. Bershadsky, A. Johansen, T. Pantev, V. Sadov,
``On Four-Dimensional Compactifications of F-Theory'',
Nucl.Phys. {\bf B505} (1997) 165, hep-th/9701165.}%
\nref\F{P. Mayr,
``Mirror Symmetry, N=1 Superpotentials and Tensionless
Strings on Calabi-Yau Four-Folds'',
Nucl.Phys. {\bf B494} (1997) 489, hep-th/9610162.}%
\nref\KLRY{A. Klemm, B. Lian, S.-S. Roan, S.-T. Yau,
``Calabi-Yau fourfolds for M- and F-Theory compactifications'',
hep-th/9701023.}%
\refs{\BJP,\F,\KLRY}.
However, geometric engineering methods involve specific local
(non-compact) degenerations which are effectively described by
field theories. Although well understood for three-folds,
the analysis is harder
in four-fold context. An important simplification would be
achieved if the relevant constructions could be carried out in
the framework of toric geometry similarly to
\ref\KMV{S. Katz, P. Mayr, C. Vafa,
``Mirror symmetry and Exact Solution of 4D N=2 Gauge Theories I'',
hep-th/9706110.}.
This has been realized for two dimensional gauge theories with
four supercharges in
\ref\L{W. Lerche, ``Fayet-Iliopoulos Potentials from Four-Folds'',
hep-th/9709146.}.
The corresponding toric model describes a resolved ADE singularity
fibered over $P^2$. In order to obtain gauge theories with
fundamental quark/anti-quark pairs, the toric diagram must be
deformed by adding extra vertices as in \KMV. These deformations
would correspond to collisions of singularities which are known
to yield matter hypermultiplets.
It turns out that in toric setup, the matter
multiplets are fibered
over rational curves, therefore they do not actually give rise
to quark multiplets in the low energy effective action
\ref\KMP{S. Katz, D. Morrison, R. Plesser, ``Enhanced Gauge Symmetry
in Type II String Theory'', Nucl.Phys. {\bf B477} (1996) 105,
hep-th/9601108.}.
This shows that the construction
must be performed beyond toric framework so that the matter
hypermultiplets can be fibered over curves of genus one.
The relevant four-fold degenerations can be regarded as generalizations
of the three-fold models considered in
\nref\MS{D.R. Morrison, N. Seiberg, ``Extremal Transitions and
Five-Dimensional Supersymmetric Field Theories'',
Nucl.Phys. {\bf B483} (1997) 229, hep-th/hep-th/9609070.}%
\nref\IMS{K. Intriligator, D.R. Morrison, N. Seiberg,
``Five-Dimensional Supersymmetric Gauge Theories and Degenerations of
Calabi-Yau Spaces'', Nucl.Phys. {\bf B497} (1997) 56,
hep-th/9702198.}%
\refs{\MS,\IMS}.

A very important dynamical aspect of gauge theories with four
supercharges is the generation of a non-perturbative superpotential
which can be exactly determined
\nref\S{N. Seiberg, ``Exact Results on the Space of Vacua of
Four Dimensional SUSY Gauge Theories'', Phys. Rev. {\bf D49}
(1994) 6857, hep-th/9402044 ; K. Intriligator, R.G. Leigh and N. Seiberg,
``Exact Superpotentials in Four Dimensions'', Phys. Rev. {\bf D50}
(1994) 1092, hep-th/9403198.}%
\refs{\S}.
In M theory, this effect is
dynamically generated by five-brane instantons wrapping compact
divisors of unit arithmetic genus
\ref\Ws{E. Witten, "Non-Perturbative
Superpotential in String Theory", Nucl.Phys. {\bf B474}
(1996) 343, hep-th/9604030.}.
The arithmetic genus constraint is closely related to the
anomalous $U(1)_R$ superselection rules governing the superpotential
in gauge theories. In the case studied in \KVa, this leads to the
expected field theory results.

In theories with fundamental matter the superpotential is affected
by quark zero modes {\BHO,\AISS}. This leads to a splitting phenomenon
dividing the Coulomb branch into different subwedges characterized by 
different instanton factors. At geometric
level, these can be understood as subcones of the extended relative
\ka\ cone as in \refs{\MS,\IMS}. The derivation of the superpotential
leads to an apparent paradox since the arithmetic genera of the
compact divisors are stable under the flavor deformations.
It turns out that the resolution involves certain subtleties
related to the dynamics of the anti-symmetric tensor field
in the five-brane world-volume. A careful analysis shows that the
anomaly controlling the superpotential can receive perturbative
contributions from the chiral two-form in the $(2,0)$ tensor
multiplet. This phenomenon has been anticipated in
\ref\Wfb{E. Witten, ``Five-Brane effective action in M theory'',
J.Geom.Phys. {\bf 22} (1997) 103, hep-th/9610234.}.

The organization of the paper is as follows: section 2 contains an
outline of the geometric construction avoiding technical details.
In section 3, we consider the quantum moduli spaces and the
superpotential from geometric point of view including a discussion of
the five-brane effective action. Section 4 provides a
more rigorous treatment of four-fold degenerations with an accent
on \ka\ phases and flops interconnecting them.  We conclude in section
5 with directions for future work.

\newsec{Outline of the basic construction}

We start with pure $SU(N)$ gauge theory on $R^3\times S^1_R$
engineered in \KVa. The three dimensional case is recovered in
the limit $R\rightarrow 0$. These theories can be found
in F-theory compactifications
on elliptic Calabi-Yau four-folds plus a circle of radius $R$.
This can be equivalently seen as a M theory compactification on an
elliptic four-fold with the size of the elliptic fiber
$\sim{1\over R}$.
The relevant degeneration consists of an $A_{N-1}$ elliptic
curve fibered over a two dimensional component $B$
of the discriminant
locus. The latter must be rational and can be taken either
$P^2$ or a Hirzebruch surface $F_n$.
Resolution of the singularity produces
a chain of $N$ $P^1$ components fibered over $B$ which intersect
according to the affine $A_{N-1}$ Dynkin diagram. As noted in \KVa,\
each $P^1$ fibered over $B$ defines a rational threefold divisor
$D_i$, $i=0\ldots N$. The divisors intersect pairwise along common
sections isomorphic to $B$. We denote the divisor corresponding to the
affine node by $D_0$. Note that $D_0$ grows to infinite size in the
limit $R\rightarrow 0$. Therefore, in three dimensions we are left only
with an ordinary Dynkin diagram as expected.
The bare gauge coupling constant is given by
\eqn\gcoupl{
{1\over g^2}=RV_B}
where $V_B$ is the base volume. The real \ka\ moduli
of the $P^1$ components combine with the periods
of the magnetic six-form of the eleven-dimensional
supergravity\foot{From gauge theoretic point of view, these are
compact scalars dual to the low energy photons along the
Coulomb branch.}
\eqn\periods{\int_{D_i}C^{(6)}}
to yield $N-1$ complex chiral multiplets $\Phi$.
These parameterize the complex one dimensional Coulomb branch
of the theory. We can introduce alternative coordinates on the
Coulomb branch
\eqn\altcoord{
A_i=\Phi_{i}-\Phi_{i-1}}
where $\Phi_0=\Phi_N=0$ by convention.
Then the Coulomb branch is identified with the standard
affine Weyl chamber
\eqn\Wchamber{
a_1>a_2>\ldots >a_N}
where $a_i$ denotes the real part of the complex scalar field
in the chiral multiplet $A_i$.

There is a non-perturbative superpotential generated by Euclidean
five-brane instantons wrapping the six-cycles $D_i$ \KVa.
The relevant instanton factors are given by
\eqn\instfact{
Y_i=\hbox{exp}\left({1\over g^2}\phi_i\right),\qquad
Y=\hbox{exp}\left({1\over g^2}\sum_{i=1}^{N-1}\phi_i\right).}
Note that only $Y$ is well defined throughout the entire Coulomb
branch due to the splitting phenomenon discussed in \AISS.
The superpotential reads
\eqn\nonpert{
W=\sum_{i=1}^{N-1}Y_i^{-1}+\gamma Y}
where
\eqn\qcdscale{
\gamma=\hbox{exp}\left(-{1\over Rg^2}\right)}
is related to the four dimensional QCD scale.

This construction can be generalized to arbitrary gauge groups
\nref\V{C. Vafa, ``On $N=1$ Yang-Mills in Four Dimensions'',
hep-th/9801139.}%
\refs{\KVa,\V}. The resolution of the singularity yields the exceptional
locus
\eqn\exloc{
\sum_{i=0}^rn_iD_i}
where $r$ is the rank of the gauge group and $n_i$ are the Dynkin
numbers of the affine diagram. The global instanton factor is then
defined by
\eqn\glb{
Y=\hbox{exp}\left({1\over g^2}\sum_{i=1}^{N-1}n_i\phi_i\right).}

The main question addressed in the present paper is the generalization
to gauge theories with matter multiplets.
Matter fields are well understood in four
and five dimensional field theories engineered from Calabi-Yau
threefolds \refs{\KMV,\MS,\IMS}
As a basic principle, quark multiplets generally arise from
collision of singularities
\nref\RM{R. Miranda, ``Smooth Models for Elliptic Threefolds'', in
R. Friedman and D.R. Morrison, editors, ``The Birational Geometry of
Degenerations'', Birkhauser, 1983.}%
\nref\BIK{M. Bershadsky, K. Intriligator, S. Kachru, D.R. Morrison,
V. Sadov, C. Vafa, ``Geometric Singularities and Enhanced Gauge
Symmetries'', Nucl.Phys. {\bf B481} (1996) 215, hep-th/96050200.}%
\nref\AG{P.S. Aspinwall, M. Gross, ``The $SO(32)$ Heterotic String
on a K3 Surface'',  Phys.Lett. {\bf B387} (1996) 735, hep-th/9605131.}%
\nref\KVb{S. Katz, C. Vafa, ``Matter From Geometry, Nucl.Phys.
{\bf B497} (1997) 146, hep-th/9606086.}%
\nref\BJ{M. Bershadsky, A. Johansen, "Colliding Singularities in
F-theory and Phase Transitions", Nucl.Phys. {\bf B489} (1997) 122,
hep-th/9610111.}%
\nref\A{P.S. Aspinwall, ``Point-Like Instantons and the
$Spin(32)/Z_2$ Heterotic String'', Nucl.Phys. {\bf B496} (1997) 149,
hep-th/9612108.}%
\refs{\RM,\BIK,\AG,\KVb,\BJ,\A}.
As opposed to three-folds, the four-fold case
relevant here is less studied. We outline the idea
of the construction, postponing the technical details for
latter sections.

Consider two $A_{N-1}$ and $A_{N_f-1}$ singularities fibered
over two components of the discriminant locus $B$, $B_f$
intersecting transversely along an elliptic curve\foot{We thank Cumrun
Vafa for explaining this construction to us.} $\Sigma$.
We take $B\simeq B_f\simeq P^2$ for concreteness.
The collision results in a non-abelian conifold singularity
\ref\HOV{K. Hori, H. Ooguri, C. Vafa, ``Non-Abelian Conifold 
Transitions and N=4 Dualities in Three Dimensions'', 
Nucl.Phys. {\bf B504} (1997) 147, hep-th/9705220.}
fibered over $\Sigma$, as detailed in appendix C. 
Resolution of singularities away from the collision locus
gives an $A_{N-1}$ tree fibered over $B$ and an $A_{N_f-1}$
tree fibered over $B_f$. 
Over $\Sigma$ , we obtain a chain of
rational components of length $N+N_f-1$ (see
\RM\ and appendix C). Note that this includes an extra
$P^1$ component arising from the resolution of a conifold
singularity.
This configuration is actually expected to yield
$SU(N)\times SU(N_f)$ gauge theory with bifundamental matter.
In order to obtain $SU(N)$ gauge theory with $N_f$ fundamental
quarks, the dynamics of $SU(N_f)$ must be weakened by sending the
size of $B_f$ to infinity.
The three dimensional quark multiplets can be thought as 
five dimensional hypermultiplets reduced on $\Sigma$. 

The resulting picture consists of a chain of $N-1$ divisors
$D_1\ldots D_{N-1}$ as in the pure gauge theory case.
As shown in section 4
and appendix C, there are different possible resolutions of the
enhanced singularity over the collision locus. These lead to
different \ka\ phases with well defined gauge theory
interpretation. For simplicity, we consider the case $N_f=1$.
The singularity can be resolved in such a way that the $l$-th
divisor $D_l$ develops a reducible fiber with two rational components
intersecting transversely over $\Sigma$.
In fact, it turns out that the divisor
$D_l$ undergoes an {\it embedded blow-up} along $\Sigma$.
It can be shown that the components of the reducible fiber
define the weights of the fundamental matter. Note that the 
resulting divisor is nonsingular and birationally equivalent 
to the rationally ruled threefold $D_l$. 

As explained in
\refs{\BHO,\AISS},
a $d=3$ $N=2$ quark multiplet
can have both complex and real mass parameters. The real mass
parameters are geometrically realized as \ka\ moduli associated to
the exceptional components defining the matter representations.
In the particular case considered above, we have a single exceptional
component whose size defines the quark real mass $m$.
As explained in \refs{\MS,\IMS},
the value of $m$ specifies the particular
\ka\ phase (subcone) of the resolution. The phase in which the
$l$-th divisor acquires a double fiber corresponds to
\eqn\kaphase{
a_1>\ldots>a_{l}>m>a_{l+1}>\ldots a_N.}
Note that these regions can be identified with the subwedges
of the Coulomb branch in gauge theory. The variation of $m$ leads
to geometrical phase transitions. The exceptional component is flopped
from one divisor in the chain to an adjoint one. It is very suggestive
to compare the geometrical flops with the brane picture as in
Fig.1.

\ifig\engZ{The flop of the exceptional fiber component between two
adjoint divisors is the geometrical counterpart of moving a flavor
D5-brane in the brane construction.}{\epsfxsize3.5in\epsfbox{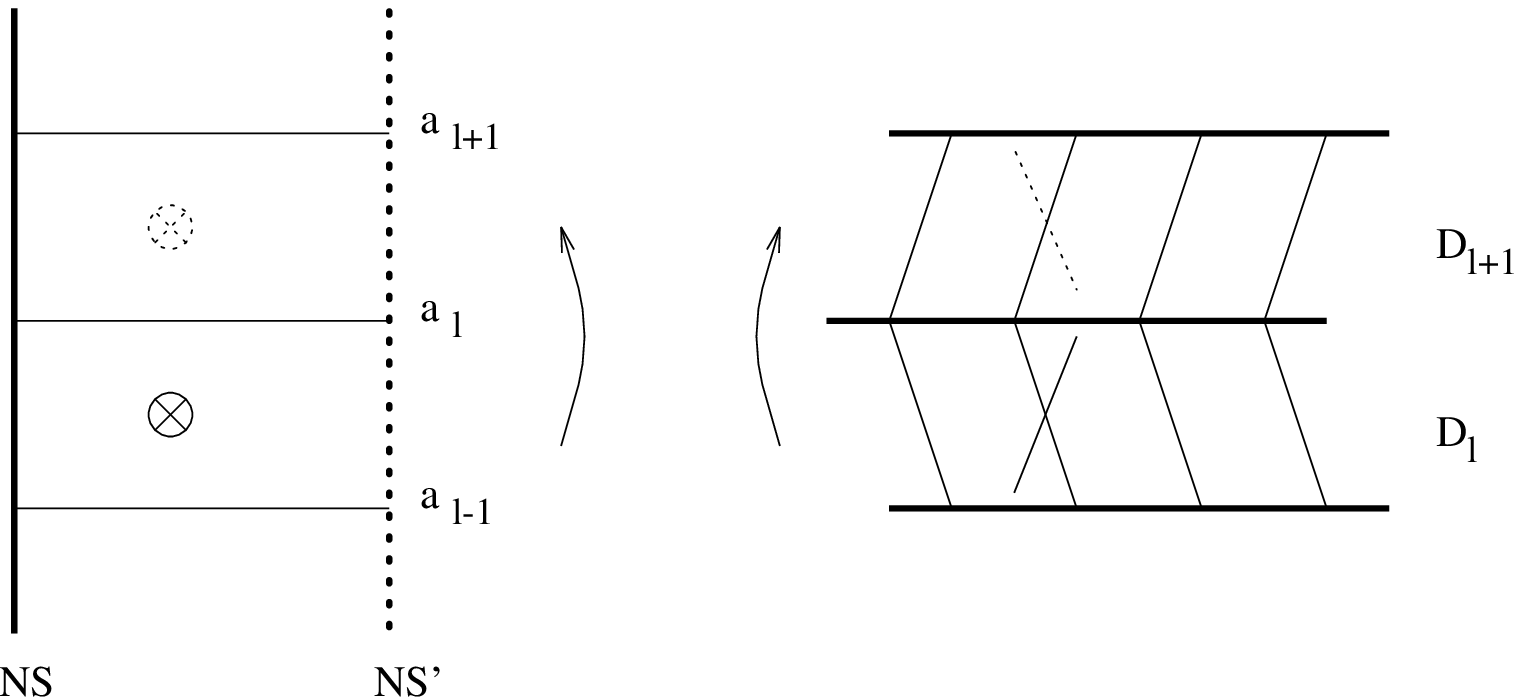}}

\newsec{Quantum Moduli Spaces and Superpotentials}

The main dynamical test of the geometric construction is to
check the non-perturbative superpotential against the gauge theory
expectations. According to \refs{\BHO,\AISS}, in the subwedge
\kaphase,\ the $l$-th instanton should be lifted 
as an effect of the quark fermion zero modes. In fact, if we have
$N_f$ flavors whose bare real masses satisfy \kaphase,\ the number
of zero modes (weighted by $U(1)_R$ charge) of the five-brane
instanton should be $1-N_f$. As we will see shortly, it turns out
that understanding this effect in geometry involves certain
subtleties.

Recall \Ws\ that generation of a superpotential
is controlled by the perturbative anomaly of
a certain $U(1)$ symmetry of the five-brane effective action.
The symmetry is generated by rotations $z\rightarrow e^{i\theta}z$
in the normal bundle $N_{D/X}$ of the six-cycle within the Calabi-Yau
space $X$. Note that the Calabi-Yau condition implies that the
normal bundle is isomorphic to the canonical bundle of the divisor
$N_{D/X}\simeq K_D$. 
For pure gauge theories,
it can be identified with the $U(1)_R$ symmetry of the gauge theory
\nref\GH{C. Gomez, R. Hernandez, ``M and F-Theory Instantons,
$N=1$ Supersymmetry and Fractional Topological Charges'',
Int.J.Mod.Phys. {\bf A12} (1997) 5141, hep-th/9701150.}%
\nref\Gomez{C. Gomez, ``Elliptic Singularities, $\theta$-Puzzle
and Domain Walls'', hep-th/9711074.}%
\refs{\GH,\Gomez}.
In general, the five-branes fermions must are twisted by
$N_{D/X}^{1/2}$. Therefore, they are sections of 
\eqn\twistsect{
{\Omega_D}^{0,0}_{-1/2}\oplus {\Omega_D}^{0,2}_{-1/2}
\oplus {K_D}_{1/2}\oplus {\Omega_D}^{0,2}\otimes {K_D}_{1/2}}
where the lower index represents $W$ charge. 
The total charge violation 
is given by the arithmetic genus of the divisor
\eqn\anomaly{
\Delta W=\chi({\co}_D)=\sum_{i=0}^3(-1)^ih^{0,i}(D).}
A superpotential is generated if
\eqn\gencond{
\chi({\co}_D)=1,\qquad h^{0,i}(D)=0,\ i=1,2,3.}

In the present situation, this leads to a puzzle since the conditions
\gencond\ are stable under blow-up. The relevant
holomorphic Hodge numbers do not change when the divisor acquires
a double fiber over the collision locus.  Therefore, it seems that
the flavor divisor $D_l$ still contributes to the superpotential,
contradicting the gauge theory results. The resolution of this puzzle
follows from a careful analysis of the effective action of the
five-brane wrapped on the blown-up divisor. There is a problem
here\foot{We thank E. Witten for explaining this to us.}
related to the quantization of the field strength $G$ of the 
supergravity three-form. As shown in 
\ref\Wflux{E. Witten, ``On Flux Quantization in M-Theory and The
Effective Action'', J.Geom.Phys. {\bf 22} (1997) 1hep-th/9609122.}, 
the correct quantization 
of $G$ in a curved background is 
\eqn\fluxq{
\int_{\Gamma} \left(G-{\lambda\over 2}\right)=0}
where $\lambda=p_1/2$ is half the Pontrjagin class of the
compactification manifold and $\Gamma$ is an arbitrary four-cycle. 
If $\lambda$ is odd, the quantization 
condition \fluxq\ requires a half-integral flux on the four-cycles 
of the manifold. Since $G$ is related to the five-brane field strength
$H$ by $G=dH$, this would prevent the wrapping of the five-brane. 
At the same time, a non-trivial $G$-flux would induce three
dimensional Chern-Simons terms which  would lift the gauge theory 
Coulomb branch \AISS. In the present situation, it can be checked
that the restriction of $\lambda$ to the blown-up divisor is even
(appendix D) avoiding these effects. 

We consider first the case when the rationally ruled divisor $D$ is
blown-up once along the elliptic curve $\Sigma$ in the base. Let
$\tilde D$ denote the resulting threefold. To fix notation, let 
$\epsilon$ denote the class of the $P^1$ fiber and $\gamma^0,\gamma^1$
denote the components of the reducible fiber over 
the curve $\Sigma$ satisfying 
\eqn\doublerel{
\gamma^0+\gamma^1= \epsilon.}
Note that $\gamma^1$ is the class of the exceptional $P^1$ while 
$\gamma^0$ is the proper transform of the original fiber. 
It will be shown in the next section that this 
geometry encodes the fundamental weights of the quarks. For now, 
we focus on the one-loop effective action of a five-brane wrapped 
on $\tilde D$. 

As pointed out in 
\nref\og{O. Ganor, ``A Note on Zeroes of Superpotentials in
F-Theory'', Nucl.Phys. {\bf B499} (1997) 55, hep-th/9612077.}%
\refs{\Ws,\og},
the non-perturbative  superpotential is of the form 
\eqn\fbsuper{
\int d^2\theta e^{-(V_D+i\phi_D)}f(m_a).}
Here $V_D+i\phi_D$ is the classical instanton action and $f(m_a)$ 
is a holomorphic function of the moduli present in the problem 
obtained by integrating out the five-brane degrees of freedom at one
loop. Since it is a determinant of massive modes, the function
$f(m_a)$ does not vanish anywhere on the moduli space in generic
situations. However, it has been shown in \Wfb\ that 
cancelations can occur\foot{We thank E. Witten for explanations and
very helpful discussions on these points.} if the six-cycle has $h_3\neq 0$. 

In the present case, the blown-up divisor $\tilde D$ has non-trivial 
homology classes in degree three arising form the exceptional $P^1$ 
curve ruled above non-trivial one-cycles of $\Sigma$. If $a,b$ denote 
a canonical set of generators of $H_1(\Sigma,Z)$ then $a\times
\gamma^1, b\times \gamma^1$ define a set of generators of 
$H_3(\tilde D,Z)$. In fact the blown-up divisor $\tilde D$ has Hodge 
numbers 
\eqn\bluphodge{
h^{0,1}(\tilde D)=h^{0,3}(\tilde D)=0,\qquad 
h^{1,2}(\tilde D)=h^{2,1}(\tilde D)=1.}
Therefore $\tilde D$ has a non-trivial intermediate Jacobian 
$J(\tilde D)=H^{1,2}(\tilde D)/H^3(\tilde D,Z)$ which is a 
principally polarized 
abelian variety 
\ref\cg{C.H. Clemens, P. Griffiths, ``The Intermediate Jacobian 
of Cubic Threefolds'', Ann. Math. {\bf 95} (1972) 281.}.
Moreover, it has been proved in \cg\ that the intermediate Jacobian
of $\tilde D$ is isomorphic (as a polarized abelian variety) to the
Jacobian of the elliptic curve
$J(\Sigma)=H^{0,1}(\Sigma)/H^1(\Sigma,Z)$. 
Note that a point 
$C\in J(\tilde D)$ parameterizes the values of the supergravity three-form on
three-cycles $a,b\times \gamma^1$ in $H_3(X,Z)$. The two real periods
combine in a complex field which is the lowest component of a chiral
multiplet \Ws. 
Under the isomorphism to $J(\Sigma)$,
$C$ can be equivalently seen as a background $U(1)$ flat
connection on the torus

The analysis of \Wfb\ shows that in this case the partition function 
of the chiral two-form in the $(2,0)$ tensor multiplet is determined
by a complex holomorphic line bundle $\cal L$ over the intermediate 
Jacobian $J(\tilde D)$. The line bundle $\cal L$ is determined up to 
translation
on $J(\tilde D)$ by the Chern class, which is equal to the polarization
$c_1(\cal L)=\omega$. The holomorphic structure can be described 
\Wfb\ in terms of a $U(1)$ connection $B$ with curvature
$F=2\pi\omega$. In order to completely determine the line bundle 
one has to specify a collection of phases $H(C)$ 
corresponding to the holonomy of $B$ around closed loops in the 
Jacobian associated to lattice vectors $C\in H^3(\tilde D,Z)$. 
In M-theory context, the phases $H(C)$ are determined by the
eleven-dimensional Chern-Simons interaction (including the
gravitational corrections) \Wfb. A precise computation is very hard to
perform since the divisor $\tilde D$ is not spin and the normal bundle
is not trivial. Alternatively, $\cal L$ can be described as follows.
The Jacobian $J(\Sigma)$ has a
distinguished holomorphic line bundle ${\cal L}_0$ associated to the 
$\Theta$ divisor. The line bundle $\cal L$ can be described as the
pull back of ${\cal L}_0$ by the map $T\rightarrow T\otimes S$ 
where $S$ is a fixed flat line bundle on $\Sigma$. In most physical
applications, $S$ is determined by a fixed spin structure on $\Sigma$ 
such that $\cal L$ is the holomorphic determinant bundle of the 
corresponding Dirac operator. In the present case, $S$ is in principle
uniquely determined by M-theory data, but it is very hard to make
explicit. 
However, this degree of accuracy suffices for our purposes. 
Since 
$c_1(\cal L)=\omega$, a simple application of the index theorem \Wfb\
shows that $\cal L$ has a unique holomorphic section up to
multiplication by a constant factor. This is the partition function 
of the chiral two-form $\theta(C)$ which depends holomorphically on
the three-form periods $C$. 

Now we can collect all the pieces and resolve our puzzle. Since the
line bundle $\cal L$ is equivalent to ${\cal L}_0$ up to translation, 
the holomorphic section $\theta(C)$ vanishes at precisely one point $C^0$
which is the translate of the $\Theta$ divisor. 
Therefore the superpotential cancels for precisely one special value 
$C^0$ of the three-from moduli. 
Clearly, the  cancelation does not take place for generic values of
$C$. This 
property identifies the periods of the three-form as complex mass
parameters in the gauge theory \BHO. More precisely the complex masses
of the quarks are given $m_c=C-C^0$. For $C\neq C^0$, $m_c\neq 0$ 
and the superpotential does not vanish in any subwedge of the Coulomb
branch. Note that the line bundle $\cal L$ and the values $C^0$ 
can depend a priori on the 
metric on the Calabi-Yau space \Wfb. However, this dependence 
is constrained by holomorphy. Since the line bundle can depend only on 
the real K\"ahler parameters by construction, it follows that it is 
actually independent of the complex K\"ahler moduli. Therefore 
the values $C^0$ are well determined once the complex structure is 
fixed. 

The discussion can be extended to an arbitrary number of flavors. 
Let $\tilde D$ denote the divisor $D$ blown-up successively $N_f$ times
along the elliptic curve $\Sigma$. Let $\gamma^1\ldots \gamma^{N_f}$ 
denote the exceptional $P^1$ components and $\gamma^0$ denote the
proper transform of the original fiber over $\Sigma$. Then, as shown in
section 4, the curve classes corresponding to
fundamental weights are given by 
\eqn\multinters{
\sigma^\alpha=\gamma^0+\ldots+\gamma^{\alpha-1},\qquad 
\alpha=1\ldots N_f.}
The three-cycles $a,b\times \sigma^\alpha$ constitute a set of
generators of $H_3(\tilde D,Z)$. Therefore $h^{1,2}(\tilde D)=N_f$ 
and the intermediate Jacobian is a $N_f$ (complex) dimensional 
polarized abelian variety. A local computation based on 
appendix B and C shows that
the three-cycles $a,b\times \sigma^\alpha$ and $a,b\times
\sigma^\beta$ do not intersect for $\alpha\neq \beta$. Therefore, 
the intermediate Jacobian is isomorphic to the $N_f$-th direct product 
$J(\Sigma)\times \ldots \times J(\Sigma)$. 
The complex line bundle is therefore 
of the form ${\cal L}\simeq \otimes_{\alpha=1}^{N_f}{\cal L}_\alpha$
where ${\cal L}_\alpha$ are line bundles as above on each factor. 
The two-from partition function reduces accordingly to a product 
of holomorphic sections $\theta_1(C_1)\ldots\theta_{N_f}(C_{N_f})$. 
The three-form periods $C_\alpha$ 
determine the complex masses of the $N_f$ flavors. The superpotential
vanishes whenever $m_c^\alpha=C_\alpha-C_\alpha^0=0$ for some $\alpha$
in agreement with gauge theory expectations \BHO.

It has been earlier noted that the classical $W$ symmetry corresponds
to the $U(1)_R$ symmetry in the low energy gauge theory. 
In the presence of quarks, the five-brane instanton is expected to
have $1-N_f$ fermion zero modes (weighted by $U(1)_R$ charge). 
Their geometric realization is rather implicit in the above
cancelation mechanism.
It is known
\ref\amv{L. Alvarez-Gaume, G. Moore, C. Vafa,'' Theta Functions,
Modular Invariance and Strings'', 
Commun. Math. Phys. {\bf 106}, (1987) 1.}
that the $\Theta$ divisor on $J(\Sigma)$ corresponds to a
distinguished
spin structure $S_0$ on $\Sigma$. The line bundle ${\cal L}_0$ 
can be represented as the determinant line bundle of the family
of twisted Dirac operators $D_{S_0}(C)$ where $C$ is a generic point 
in $J(\Sigma)$. The zero in the Riemann theta function is then 
associated to a cancelation of the determinant as a result of 
a fermionic zero mode. 
Since ${\cal L}$ is obtained from ${\cal L}_0$ by translation, 
we conclude that the zero of the section of ${\cal L}$ can also
be thought as a cancelation due to fermionic zero mode. In fact 
${\cal L}$ can be thought as a determinant of a family of
twisted Dirac operators $D_S(C)$, but the origin $D_S(0)$ being
twisted by the line bundle $S$. The cancelation occurs for the
special value of $C$ which undoes the effect of the initial twist. 

Since these fermion zero modes appear in a rather abstract manner, 
the question is how to interpret them. As shown in \Ws,\ the $W$
charge violation can be interpreted as a diffeomorphism anomaly 
which must be cancelled by a classical effect. As a result, the 
compact magnetic scalar $\phi_D$ is shifted by $\chi({\co}_D)$ 
under rotations in the normal bundle. 
In the above discussion, the cancelation of the superpotential can be 
effectively seen as an effect of chiral fermion zero modes localized
along $\Sigma$. Since the fermions are locally sections of 
${K_\Sigma}^{1/2}$ (ignoring the twist), 
 these can contribute to the $W$ charge anomaly if the
latter is localized accordingly, that is if $W$ acts trivially in
normal directions to $\Sigma$. In local coordinates, if $z$ is a 
coordinate on $\Sigma$ and $t,w$ are normal coordinates, $W$ should
act as $dz\rightarrow e^{i\theta}dz$, $dt,dw\rightarrow dt,dw$. Equivalently,
the diffeomorphisms generated by $W$ are required to preserve a normal
neighborhood of $\Sigma$. The net effect is a $N_f$ contribution to 
the $W$ charge violation, but there is a subtle question related to
the sign of this contribution. In the present conventions, the chiral
two-form in the $(2,0)$ tensor multiplet has self-dual field strength
$H$ which couples to the anti-self dual part of $C$, therefore 
the partition function is a holomorphic section over
$J(\tilde D)\simeq J(\Sigma)^{N_f}$ rather than anti-holomorphic \Wfb.  
This means that the twisted Dirac operator couples to holomorphic 
flat bundles on $\Sigma$, therefore the fermionic zero modes are 
locally sections of ${K_\Sigma}^{1/2}$ rather than
$\overline{K_\Sigma}^{1/2}$. Therefore the fermion zero modes have 
$W=1/2$. Comparing to \twistsect,\ we conclude that 
\eqn\locanom{
\Delta W=1-N_f}
for the blown-up divisor. 
This analysis suggests that in this case, the shift 
in $\phi_D$ picks up an extra $-N_f$ contribution from the diffeomorphism
anomaly of the chiral two-form. A complete derivation of this 
effect is very subtle and will not be given here. 

\subsec{Quantum moduli spaces}

We have seen above that the Coulomb branch of the three dimensional 
gauge theories can be identified with extended K\"ahler cones of 
Calabi-Yau degenerations. 
The geometric nature of Higgs branches can be similarly understood 
since the 
boundaries between \ka\ subcones correspond to points where
the quark multiplets become massless. 

Consider an $SU(N)$ gauge theory
with $N_f$ flavors in the subwedge \kaphase.
A Higgs branch emerges from the boundary of
the subcone when the flavors becomes massless $a_l=m$. Since the
corresponding exceptional $P^1$ is of zero
size, the Calabi-Yau space is singular at the origin. The expectation
values of the quark fields parameterize complex structure deformation
of the singular locus. Therefore, we obtain an extremal transition
between two branches of the Calabi-Yau moduli space generalizing
the known conifold transitions. The discussion has been so far
classical. Quantum mechanically, the Coulomb branch is complexified
by adding the six-form periods \periods.\ The emerging complex
structure branch is invariant under arbitrary shifts of the latter,
therefore, it should connect to the Coulomb branch along a complex
codimension one locus.
Thus we have recovered the quantum structure of the moduli space
described in \refs{\BHO,\AISS}, from geometric point of view.
The effective
superpotential can be written
\eqn\effect{
W_{pert}=-N_f\left(V_+V_- \dm\right)^{1\over N_f}}
where $V_{\pm}$ are holomorphic coordinates in the two
regions of the Coulomb branch and $\dm$ is a gauge invariant
coordinate on the Higgs branch. Similar transitions and the 
associated superpotentials have been discussed from F-theory 
point of view in 
\ref\grassi{A. Grassi, ``Divisors on ellipltic Calabi-Yau 
4-folds and the suprpotential in F-theory, I'', alg-geom/9704008.}

In general, the exact dependence of the superpotential on the
Coulomb branch variables is hard to determine due to the splitting
phenomenon discussed in \AISS. Once the geometric origin of the
$U(1)_R$ charge understood, the analysis follows the same lines
as in the gauge theory.
We restrict to $SU(N)$ theory with $N_f\leq N-1$ of equal mass $m$.
As explained in appendix A, the extended \ka\ cone is divided into
$N-1$ subcones separated by $N-2$ boundaries defined by $a_l=m,
l=2\ldots N-2$. Each boundary introduces a splitting region of the Coulomb
branch which connects to the Higgs branch. This imposes certain
restrictions on the instanton factors which cannot be analytically
continued past the splitting regions. The procedure is explicitly
carried out in appendix A. 

The maximal region of the Coulomb branch unlifted
by non-perturbative effects is semiclassically \AISS
\eqn\unlifted{
a_1>a_2=\ldots =a_{N-1}=m>a_N}
It can be parameterized by the global instanton
factor $Y$.  The effective superpotential as a function of
$(Y,\dm)$ can be determined \AISS\ by integrating out all the 
other fields in (A.6)
\eqn\effectsuper{
W\sim -(N_f-N+1)\left(Y\dm\right)^{1\over N_f-N+1}.}

Note that this result can be directly determined in M theory
as an instanton effect associated to a five-brane wrapping the
divisor sum $D_1+\ldots +D_{N-1}$. This corresponds to a monopole 
of charges $(1,1\ldots 1)$ in the gauge theory which is expected 
to have $N-N_f-1$ zero modes \BHO. 
The number of zero modes cannot be determined by a direct method
because the Calabi-Yau space is singular along the sub-locus \unlifted.
Since it is independent of K\"ahler moduli, it can be computed 
for the blown-up Calabi-Yau space and then extrapolated by an 
adiabatic argument. 

The sum $D_1+\ldots +D_{N-1}$ is a singular divisor with normal 
crossings whose arithmetic genus can be computed using 
Grothendieck-Riemann-Roch theorem (appendix E)
\eqn\arthgen{
\chi({\co}_D)=1.}
This apparently leads to a disagreement with the gauge theory result. 
Note however that the $N-1$ branches of the five-brane
can be separated in the transverse noncompact directions by giving
expectation values to the scalars in the tensor multiplet. This
corresponds to separating the monopole centers of charge in the
gauge theory. Since the branches do not touch each other, each of
them can be analyzed separately giving precisely one fermion zero
mode. Therefore, the net $W$ charge violation is
\eqn\anomC{
\Delta W=N-N_f-1.}
When the $N-1$ branches come together, there are two sources of
extra fermion zero modes.
As also pointed out in \Gomez,\
there are extra fermionic zero modes of negative charge signaling
the occurrence of domain walls in four dimensions. The computation
in appendix E shows that they are associated with corrections to
the arithmetic genus coming from singular points of the sum 
divisor. At the same time there are new hypermultiplet degrees of
freedom when two five-brane branches intersect along a section. 
The origin of these hypermultiplets has been described in \Wb. 
To see that these fermions contribute to the anomaly, note that 
the zero mode problem can be analyzed by scaling the size of the base
to be much larger than the size of the fiber. In this limit the
degrees of freedom in the $(2,0)$ tensor multiplet can be effectively 
reduced along the $P^1$ fiber. This results in a twisted $N=2$
hypermultiplet localized along the base $B$. The twist is easily
derived by reducing the fermions \twistsect\
\eqn\twistbase{
{\Omega_B}^{0,0}_{-1/2}\oplus{\Omega_B}_{-1/2}^{0,2}\oplus
{\Omega_B}^{0,1}\otimes {K_B}_{1/2}.}
This shows that the number of zero modes weighted by $W$ charge is
given by $\chi({\co}_B)$. In fact we have derived a physical
realization of the spectral sequence technique used in \KVa. 
The net effect is that each hypermultiplet 
contributes $\chi({\co}_B)$ to the $W$ charge anomaly. Therefore, the 
final result is 
\eqn\totWB{
\Delta W=1+N-2-N_f=N-N_f-1.}

If $N_f=N-1$, we have $\Delta W=0$ and the expected quantum effect is
a deformation of the classical moduli space of the form
\eqn\deform{
Y{\dm}=1.} 
If $N_f=N-2$, $\Delta W=1$ and we expect a non-perturbative
superpotential $\sim{1\over Y}$. The dependence on the Higgs branch
variables can be obtained following the instanton through the
transition discussed above. Gauge invariance and holomorphy
would restrict the dependence to an analytic function of $\dm$.
However, since the origin is a singular point on the Higgs branch,
the function is actually allowed to have a pole there \BJV.
Moreover, such a pole is expected on general grounds
\ref\Kachru{S. Kachru, ``Aspects of N=1 String Dynamics'',
Nucl.Phys.Proc.Suppl. 61A (1998) 42, hep-th/9705173.}.
We conjecture\foot{It would be very interesting to
check this conjecture by explicitly evaluating the determinants
of the non-zero modes in the five-brane theory. A similar computation
for Ray-Singer analytic torsion has appeared in
\ref\cecotti{M. Bershadsky, S. Cecotti, H. Ooguri, C. Vafa,
``Kodaira-Spencer Theory of Gravity and Exact Results for
Quantum String Amplitudes'', Commun.Math.Phys. {\bf 165} (1994) 311,
hep-th/9309140.}.}
that the behavior of the
instanton generated superpotential is of the form
\eqn\pole{
W\sim{1\over Y\dm}.}
Therefore we obtain the expected gauge theory results. 

\newsec{The geometry of Calabi-Yau degenerations}

Here we describe in detail the Calabi-Yau four-fold degenerations
leading to $d=3$ $N=2$ gauge theories with matter and simply laced 
gauge groups. The geometry corresponding to non-simply laced gauge 
groups is more complicated and will not be treated here. 
The construction is a generalization of the threefold models
studied in \refs{\MS,\IMS}.

\subsec{$SU(N)$ with $N_f$ quarks}

As presented in section 2, pure $SU(N)$ gauge theory corresponds to a
chain $D_1\ldots D_{N-1}$ of rationally ruled divisors over a rational
base $B\simeq P^2$.
Let $\epsilon_1\ldots\epsilon_{N-1}$ denote the
curve classes of the rational fibers. Since the whole construction is
embedded in a Calabi-Yau space, it is easy to prove (appendix B)
that the intersection matrix
\eqn\intersA{
D_i\cdot \epsilon_k=\left\{\matrix{
-2\hfill &\qquad k=i\hfill\cr
1\hfill &\qquad |i-k|=1\hfill\cr
0\hfill &\qquad k\neq i,i\pm 1\hfill\cr}\right.}
is the negative Cartan matrix of $SU(N)$.
The weights of the fundamental matter correspond to extra curve classes
$\sigma_1^\alpha\ldots\sigma_{N}^\alpha$, $\alpha=1\ldots N_f$,
satisfying
\eqn\intersB{
D_i\cdot\sigma_k^\alpha=\left\{\matrix{
-1\hfill &\qquad i=k\hfill\cr
1\hfill &\qquad k=i+1\hfill\cr
0\hfill &\qquad \hbox{otherwise}\cr}\right.}

Real bare mass parameters correspond to extra divisors $M_\alpha$
in the ambient Calabi-Yau space \refs{\MS,\KMV}
such that
\eqn\intersC{
M_\alpha\cdot \sigma^\beta = \delta_{\alpha}^{\beta},
\qquad M_\alpha\cdot\epsilon_k=0.}
An arbitrary divisor supported on the exceptional locus can be written
\eqn\arbdiv{
D=\sum_{i=1}^{N-1}\phi_i D_i+
\sum_{\alpha=1}^{N_f}m_{\alpha}M_{\alpha}}
where $\phi_i$ are order parameters on the Coulomb branch
and $m_\alpha$ are bare masses.
Following \IMS,\ we can introduce new coordinates on the Coulomb
branch
\eqn\newcoord{
a_i=\phi_i-\phi_{i-1},\qquad i=0\ldots N}
with $\phi_0=\phi_{N}=0$.
The negative \ka\ cone is divided into subcones defined by the relations
\eqn\relA{\eqalign{
& -D\cdot\epsilon_k>0,\qquad k=1\ldots N-1\cr
& -D\cdot\sigma_k>0,\qquad k\leq l\cr
& -D\cdot\sigma_k<0,\qquad k\geq l+1.\cr}}
It follows that the extended \ka\ cone corresponds to the standard
Weyl chamber
\eqn\Wchamber{
a_1>a_2>\ldots > a_{N}.}
The subcones are subwedges of the Weyl chamber defined by
\eqn\subwA{
a_1>\ldots>a_l>m_\alpha>a_{l+1}>\ldots a_{N}.}
In geometrical terms, this means that the weight $\sigma_\alpha$
is represented by an irreducible holomorphic curve such that
\eqn\intersD{
\sigma_l\cdot D_l=-1.}
This can be realized by blowing-up the $l$-th divisor along an
elliptic curve $\Sigma$ in the base. The resulting
threefold can be viewed as a $P^1$ fibration over a two
dimensional base with a reducible double fiber over $\Sigma$.
This is the higher dimensional analogue of the
exceptional curves of the first kind introduced in \IMS.
The resulting configuration is represented
schematically in Fig.3. Bold vertical lines represent sections
i.e. two dimensional surfaces. Dotted horizontal lines represent
the $P^1$ fibers. The reducible fibers corresponding to matter
representations is represented by a continuous line. It should be
understood that they are fibered over an elliptic curve $\Sigma$
in the base.

\ifig\engA{The four-fold degeneration corresponding to $SU(N)$ theory
with one flavor.}{\epsfxsize5.0in\epsfbox{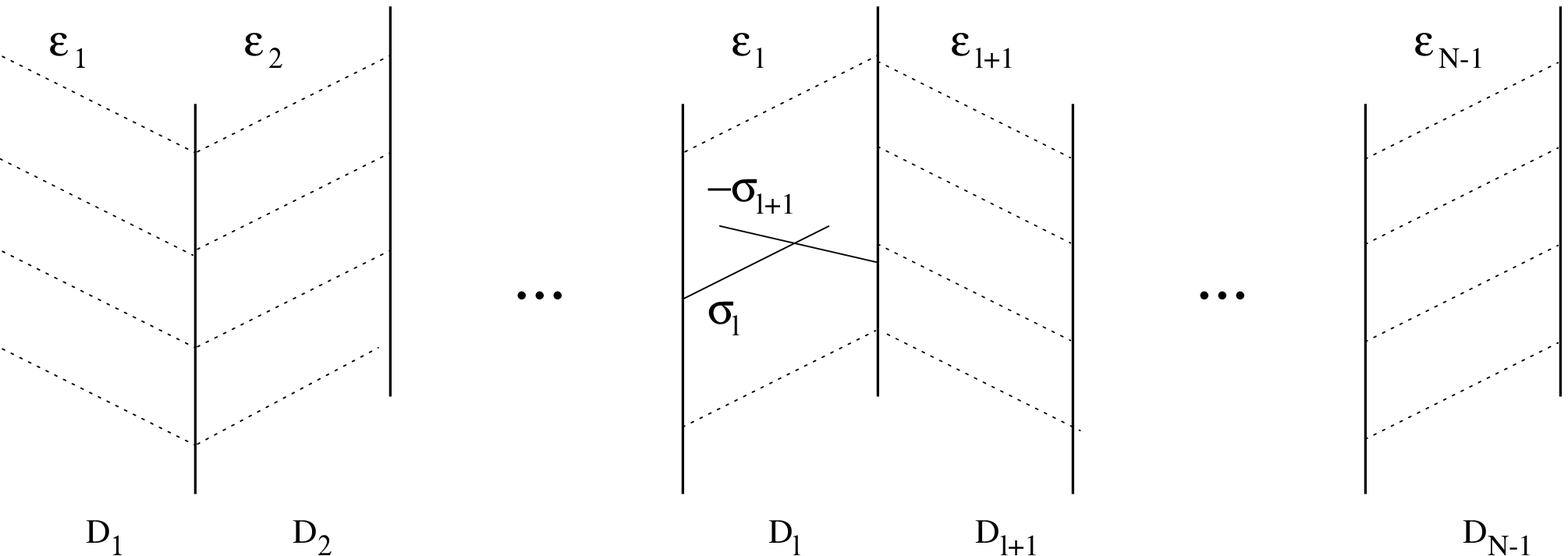}}

\noindent
The fundamental weights are represented by the connected
holomorphic curves
\eqn\weights{\eqalign{
& \sigma_k=\epsilon_k+\ldots\epsilon_{l-1}+\sigma_l,\qquad k\leq l\cr
& -\sigma_k=-\sigma_{l+1}+\epsilon_{l+1}+\ldots+\epsilon_{k-1},
\qquad k\geq l+1.\cr}}
Note that the irreducible rational curves $\sigma_l,-\sigma_{l+1}$
are the two components of the double fiber
\eqn\relB{
\sigma_l+(-\sigma_{l+1})=\epsilon_l.}

When the number of flavors is $N_f >1$, the $l$-th divisor is blown-up
$N_f$ times along $\Sigma$. This results in a reducible fiber with
$N_f+1$ components $\gamma_l^0,\ldots\gamma_l^{N_f}$ represented in
Fig.4. The intersection numbers
$\gamma_l^\alpha\cdot D_l$ are $-1,0\ldots 0,-1$ (appendix B).

\ifig\engB{The reducible fiber over the collision locus for
$N_f\geq 2$.}{\epsfxsize3.5in\epsfbox{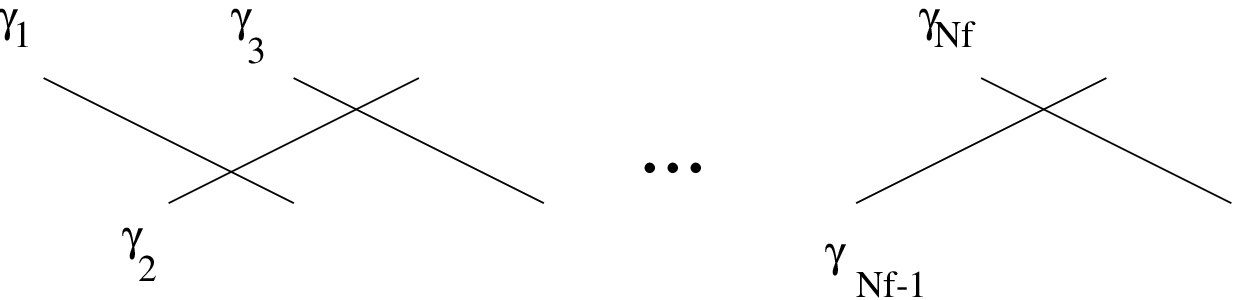}}

\noindent
The $N_f$ fundamental weights are determined by the reducible
curves
\eqn\weightsB{
\sigma_l^\alpha=\gamma_l^0+\gamma_l^1+\ldots+\gamma_l^{\alpha-1}.}

Two adjoint subcones are related by a flop of a flavor curve, moving
a fiber component from one divisor to the other. Note that this
transition passes through a singular point when the flavor component
is shrunk to zero size, i.e. when we have
\eqn\singA{
m_\alpha=a_l}
for some $l$ and $\alpha$. At this point it is
possible to move
on a different branch by a complex deformation of the singularity
rather than a \ka\ blow-up.
In gauge theory, this corresponds
classically to a transition to the Higgs branch, as explained in
section 3.

Note that the degeneration constructed above can be naturally
obtained as resolution of an $A_{N-1}+A_{N_f-1}$ collision
\refs{\RM,\BIK,\AG,\KVb,\BJ,\A}.
The two singularities are trivially fibered over two $P^2$ components
of the singular locus intersecting along the elliptic curve $\Sigma$. The
size of the flavor $P^2$ must be taken to infinity in order to
reduce $SU(N_f)$ to a global symmetry group. The divisors $D_1\ldots
D_{N}$ result from the resolution of the $A_{N-1}$ singularity.
The reducible fibers corresponding to fundamental weights arise
from the resolution of the $A_{N_f-1}$ over the collision locus,
as detailed in appendix C.

\subsec{$Spin(2N)$ with fundamental matter}

The degeneration corresponding to $Spin(2N)$  consists of $N$ divisors
$D_1\ldots D_N$ intersecting along sections as below.

\ifig\engH{The Calabi-Yau degeneration corresponding to $Spin(2N)$
with fundamental quarks.}{\epsfxsize4.5in\epsfbox{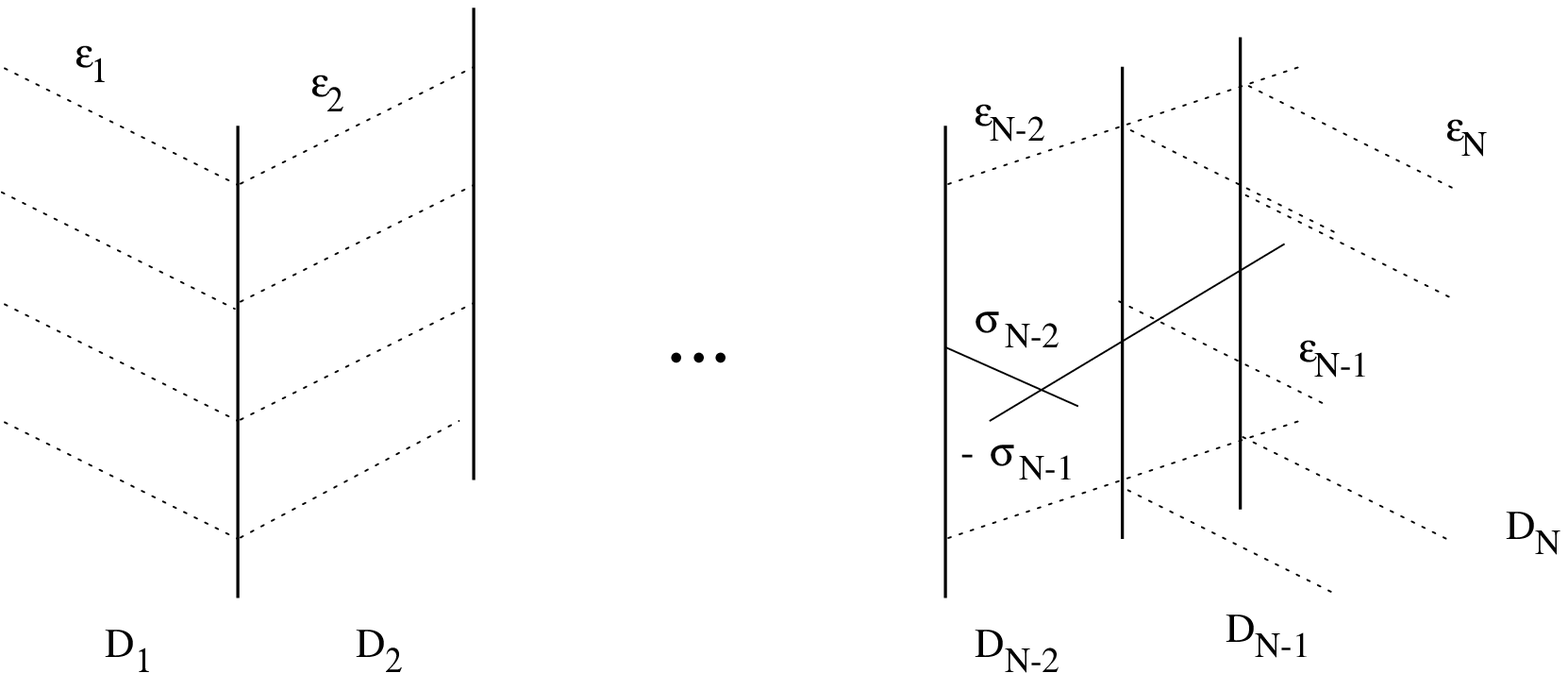}}

\noindent
If $\epsilon_1\ldots \epsilon_N$ denote the $P^1$ fiber classes,
the intersection matrix
\eqn\evspcartan{
D_i\cdot \epsilon_k=\left\{\matrix{
-2\hfill &\qquad k=i\hfill\cr
1\hfill &\qquad |i-k|=1,\ i\neq N,\ k\neq N\hfill\cr
1\hfill &\qquad (i,k)=(N-2,N)\ \hbox{or}\ (N,N-2)\cr
0\hfill &\qquad \hbox{otherwise}\hfill\cr}\right.}
is the negative $SO(2N)$ Cartan matrix.
The fundamental weights are represented by connected holomorphic
curves $\sigma_1\ldots \sigma_N$ satisfying
\eqn\evspwA{
D_i\cdot \sigma_k=\left\{\matrix{
-1\hfill & \qquad k=i \hfill\cr
1\hfill & \qquad  k=i+1\hfill\cr
-1\hfill & \qquad (i,k)=(N,N-1)\hfill\cr
0\hfill & \qquad \hbox{otherwise}\hfill\cr}\right.}
\eqn\evspwB{
\sigma_{N-1}+\sigma_N=\epsilon_N.}

An arbitrary divisor supported on the exceptional locus
can be written as
\eqn\evspdiv{D=\sum_{i=1}^{N-1}\phi_iD_i+{1\over 2}
\phi_N(D_N-D_{N-1})+mM,}
including the mass divisor $M$.
We introduce alternative Coulomb branch coordinates
\eqn\evspcoord{
a_k=\phi_k-\phi_{k-1}.}
Fundamental weights can be realized in the standard manner by
blowing-up one of the first $N-2$ divisor along an elliptic curve in
the base. If we blow-up $D_l$, the relevant curve classes are
\eqn\evfw{\eqalign{
& \sigma_k=\epsilon_k+\ldots\epsilon_{l-1}+\sigma_l,\qquad k\leq l\cr
& -\sigma_k=-\sigma_{l+1}+\epsilon_{l+1}+\ldots+\epsilon_{k-1},\qquad
l+1 \leq k\leq N-1.\cr}}
The corresponding \ka\ subcone is defined by
\eqn\evsubcone{\eqalign{
-D\cdot\sigma_k>0,\qquad k\leq l\cr
-D\cdot\sigma_k<0,\qquad k\geq l+1\cr}}
which yield
\eqn\evsubw{
a_1>\ldots>a_l>m>a_{l+1}>\ldots>a_N>0.}
For $l=N-2$, the components of the double fiber must intersect
the divisors $D_{N-1},D_N$ as in Fig.8.
As noted in $SU(N)$ case, the subcones of the \ka\ cones are related
by flops moving the singular fiber from one divisor to the other.
Since the geometry is different, the flop relating the subcones
$N-2,N$ or $N-2,N-1$ is expected to have a peculiar behavior.
Taking into account the exchange
symmetry $D_{N-1}\leftrightarrow D_N$, it suffices to treat only one
case, say $N-2,N$. We first describe the geometry which realizes the
fundamental weight $\sigma_N$ as an irreducible fiber in $D_N$.
The construction is similar to the $Spin(2N)$
threefold degeneration considered in \IMS.

The weights $\sigma_{N-1},\sigma_{N}$ must satisfy
simultaneously
\eqn\simult{\eqalign{
& \sigma_{N-1}+\sigma_N=\epsilon_N\cr
& \sigma_{N-1}=\epsilon_{N-1}+\sigma_N.\cr}}
Since $\sigma_N\cdot D_N=-1$, it follows that $\sigma_N$
must be the fiber of an exceptional divisor in $D_N$
introduced by blowing-up along a curve $\Sigma$ in the base.
Moreover, $\Sigma$ must be an elliptic curve in order
to obtain exactly one flavor. We also have $D_{N-1}\cdot
\sigma_{N}=1$, therefore the divisors $D_{N-1},D_N$ must
have nontrivial intersection. Equations \simult\ imply that
\eqn\impl{
\epsilon_N=\epsilon_{N-1}+2\sigma_N.}
Therefore, the fiber of the ruling of $D_N$ can be written as
a sum between a rational component $\delta$ and $2\sigma_N$
over $\Sigma$. The curve $\delta$ is identified with the fiber of
$D_{N-1}$ over $\Sigma$. We conclude that the divisors
$D_{N-1},D_N$ intersect along a rationally ruled surface
over $\Sigma$ with fiber $\delta\sim \epsilon_{N-1}$.
Note that $\Sigma$ is the collision locus of two sections of
$D_{N-1},D_N$ as in Fig.9.

\ifig\engI{$Spin(2N)$ with one fundamental flavor in the ``special''
subcone of the extended \ka\ cone.}{\epsfxsize4.5in\epsfbox{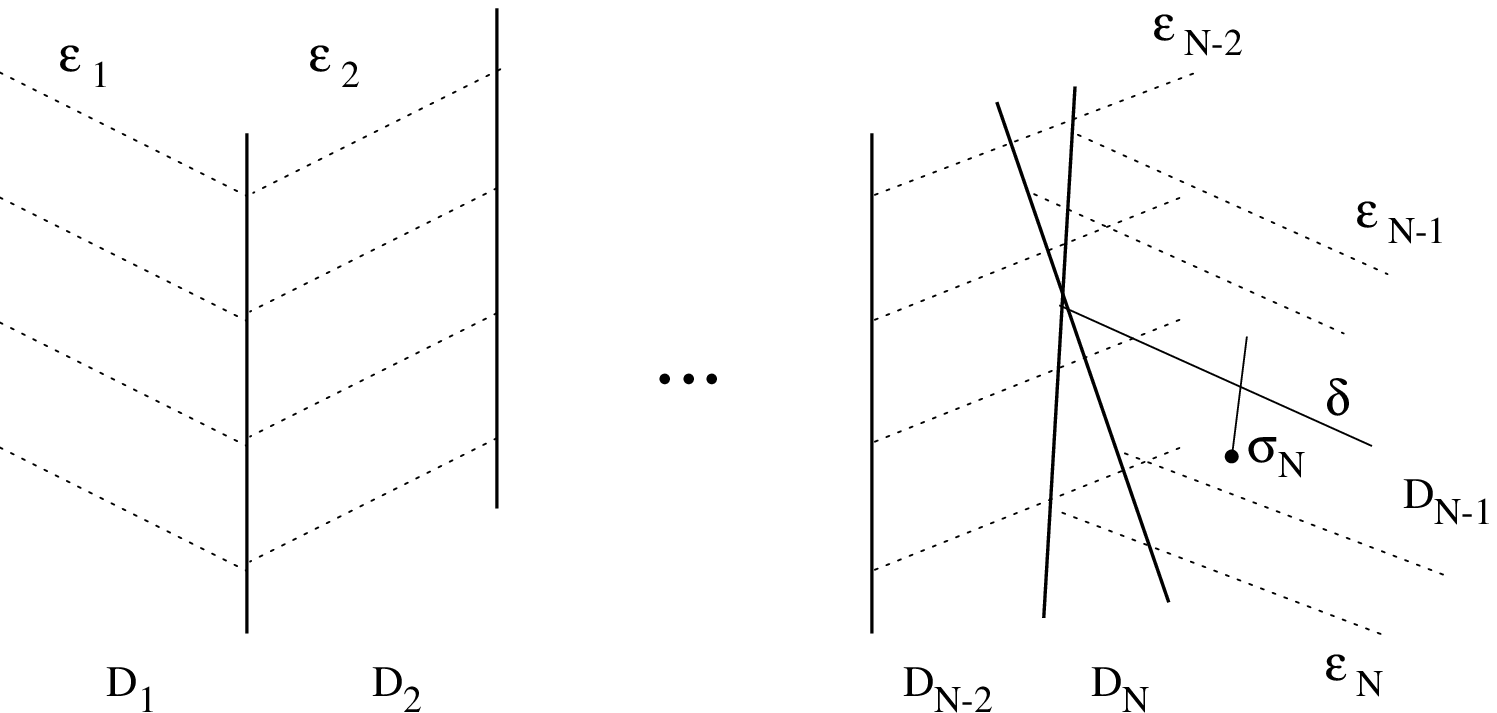}}

\noindent
Finally, the intersection number $\delta\cdot D_N$ must be $0$,
therefore the normal bundle of $\delta$ in $D_N$ must have degree
$-2$. This is similar to the situation described in \IMS. If we restrict
to a local analysis in directions normal to the curve $\Sigma$, it
follows that the curve $\sigma_N$ must pass through an ordinary
double point as in \IMS. The difference is that in the present
situation, the singularity is fibered over $\Sigma$, therefore
the divisor $D_N$ has a curve of singularities.

The flop relating the model in figure Fig.8 to that in Fig.9 can be
understood as follows. First, we contract the curve $\sigma_{N-1}$
in Fig.8, introducing a curve $\Sigma$ of ordinary double points in the
ambient
Calabi-Yau space. This results in a collision of the divisors
$D_{N-1}$, $D_{N}$
along a rationally ruled surface over the torus $\Sigma$ with fiber
$\delta$. The common section $\delta$ passes through the singular
locus of the four-fold. The singularity of the total space can be
resolved by performing an embedded blow-up of the divisor $D_N$.
Therefore $D_N$ acquires an exceptional curve of the first kind
fibered over the elliptic curve $\Sigma$. As shown above, this is
configuration cannot be embedded in a smooth Calabi-Yau unless
$D_N$ develops at the same time a double point singularity
along $\Sigma$.

The \ka\ subcone corresponding to the present construction
is defined by
\eqn\secevsubcone{
-\epsilon_k\cdot D>0,\qquad -\sigma_N\cdot D_N>0.}
Hence it is identified with the
\eqn\specevw{
a_1>a_2\ldots>a_N>m>0}
subwedge of the Coulomb branch.

\newsec{Conclusions}

To summarize our results, we have extended the methods of geometric
engineering to $N=2$ gauge theories with fundamental matter.
The construction is rather general, including
various gauge groups with distinct matter content.
Even though we mainly discuss the Coulomb branch of the field
theories, we also provide information
on the Higgs branch corresponding to complex deformations of
the singular locus. For instance, we verify that the
dynamically generated superpotentials on both
branches match the gauge-theoretic expectations.
Along with a geometric interpretation of the $U(1)_R$ charge,
the analysis involves a careful zero-mode counting of five-brane
instantons. Remarkably, the agreement with gauge theory is
based on subtle aspects of the five-brane world volume dynamics.
The flavor contribution to the $U(1)_R$ charge anomaly
is results from the effective action of the chiral two-form
in the $(2,0)$ tensor multiplet. We also derive an interesting
prediction concerning poles in the superpotential at the
origin of the Higgs branch. It would be very interesting
to explicitly check this prediction by computing the
determinants of non-zero modes in the five-brane effective action.
Presumably, the dependence on the complex structure moduli is
similar to the dependence of Ray-Singer analytic torsion
on the gauge field moduli.

Needless to mention, this article poses a number of questions
which we hope to tackle in the future. Aside the superpotential term,
there is less understood effective \ka\ potential of the gauge
theory.  Apparently the geometric considerations
are powerless against this term, since it is not protected from
quantum corrections. At a closer look one can find that
holomorphic topological
data somehow enters the \ka\ potential \L. This fact is not surprising since
the Coulomb moduli space should be {\it an analog of the special \ka\
manifold} for higher-dimensional Calabi-Yau spaces
\ref\GMP{B.R.Greene,
D.R.Morrison, M.R.Plesser, "Mirror Manifolds in Higher Dimension", Commun.
Math.  Phys.  {\bf 173} (1995) 559, hep-th/9402119.}
and is expected to serve some constraints
on (if not totally define) the \ka\ potential. Also at present we lack
understanding of what happens when the number of quarks becomes large and
equal to special values, e.g. $2N_c$, $3N_c$.  This question touches
non-abelian duality whose purely geometric nature remains elusive, even though
a great step in this direction was done in \OV.

\bigskip
\centerline{\bf Acknowledgments}

We are very grateful to Cumrun Vafa and Edward Witten for
collaboration and
valuable suggestions and to Liviu Nicolaescu for mathematical
assistance.
We would also like to thank Ofer Aharony, Tom Banks,
Jan de Boer, Michael Douglas, Rami Entin, Ori Ganor, Jaume Gomis,
Brian Greene, Barak Kol, Wolfgang Lerche, Nathan Seiberg
and Piljin Yi for very helpful discussions and
correspondence.
S.G also would like
to thank Laboratoire de Physique Th\'eorique et Hautes Energies
where a part of this work was done,
and especially L. Baulieu for kind hospitality and the
support of the CNRS grant.

The research of S. G. was supported in part by Merit
Fellowship in Natural Sciences
and Mathematics, grant RFBR-96-15-96939 and CNRS Foundation.

\bigskip

\appendix{A}{The superpotential for $SU(N)$ theory with $N_f$ flavors}

As stated in the text, we consider only the case when the flavors have
equal mass $m$. The extended \ka\ cone is divided in $N-1$ subcones
$K_1\ldots K_{N-1}$ characterized by
\eqn\ksubcones{
a_1>\ldots >a_l>m>a_{l+1}>\ldots a_{N}.}
Let $Y_{1,l}\ldots Y_{N-1,l}$ denote the corresponding instanton
factors satisfying $Y=Y_{1,l}\ldots Y_{N-1,l}$. Note that
$Y_{i,l}, i\neq l$ have $W$ charge $-1$ while $Y_{l,l}$ has $W$ charge
$N_f-1$. Semiclassically, we have $Y_{i,l}\sim
e^{(A_i-A_{i+1})/g^2}$. These relations can be inverted as follows
\eqn\invert{\eqalign{
& e^{A_{i}/g^2}=
\left({Y_{i,l}^{N-i}Y_{i+1,l}^{N-i-1}\ldots Y_{N-1,l}\over
Y_{1,l}Y_{2,l}^2\ldots Y_{i-1,l}^{i-1}}\right)^{1\over N}.\cr}}
Note that each boundary between subcones of the form $a_l=m$ introduces a
splitting in the Coulomb branch. The perturbative 
superpotential in that region is
is of the form
\eqn\splitting{
W\sim -N_f\left(V_{+l}V_{-l}\dm\right)^{1\over N_f}.}
where $V_{\pm l}$ are Coulomb branch variables given semiclassically
by $V_{\pm l}\sim e^{\pm \alpha A_{l}/g^2}$ where $\alpha$ is
determined such that \splitting\ has the correct $U(1)_R$ charge.
Using \invert,\ we obtain
\eqn\Vfactors{
V_{+l}V_{-l}=\left({Y_{l,l}^{N-l}Y_{l+1,l}^{N-l-1}\ldots Y_{N-1,l}\over
Y_{1,l}Y_{2,l}^2\ldots Y_{l-1,l}^{l-1}}\right)^{\alpha\over N}
\left({Y_{l,l-1}^{N-l}Y_{l+1,l-1}^{N-l-1}\ldots Y_{N-1,l-1}\over
Y_{1,l-1}Y_{2,l-1}^2\ldots Y_{l-1,l-1}^{l-1}}\right)^{-{\alpha\over
N}}.}
The exponent $\alpha$ is determined such that the $U(1)_R$ charge of
$\left(V_{+l}V_{-l}\right)^{1\over N_f}$ is $2$. We find
\eqn\exponent{
\alpha={N-1\over N}.}
Introducing the global instanton factor $Y$, \Vfactors\ becomes
\eqn\VfactorsB{\eqalign{
V_{+l}V_{-l}=&
\left({Y^{N-1}\over Y_{1,l}^{N-l+1}Y_{1,l-1}^{l-2}
Y_{2,l}^{N-l+2}Y_{2,l-1}^{l-3}\ldots
Y_{l-1,l}^{N-1}Y_{l,l-1}^{N-1}}\right)^{1\over N-1}\cr
&\left({1\over Y_{l+1,l}Y_{l+1,l-1}^{N-2}\ldots Y_{N-1,l}^{N-l-1}Y_{N-1,l-1}^l}
\right)^{1\over N-1}.\cr}}
The instanton factors in $Y_{1,l-1}\ldots Y_{l-2,l-1}$
can be analytically continued across the boundary since they are not
adjoint to the splitting region. In fact we can choose independent
coordinates on the Coulomb branch
$Y_{1,N-1}\ldots Y_{N-2,N-1}$, $Y_{21}\ldots Y_{N-1,N-2}$ and $Y$.
Note that there are exactly $2N-3$ as argued in \AISS.
Then
\eqn\VfactorsC{
V_{+l}V_{-l}={Y\over Y_{1,N-1}Y_{2,N-1}\ldots Y_{l-1,N-1}
Y_{l,l-1}Y_{l+1,l}\ldots Y_{N-1,N-2}}.}
We finally obtain the complete superpotential
\eqn\complsuper{\eqalign{
W=& -N_f\sum_{l=1}^{N_1}\left({Y\dm\over Y_{1,N-1}Y_{2,N-1}\ldots Y_{l-1,N-1}
Y_{l,l-1}Y_{l+1,l}\ldots Y_{N-1,N-2}}\right)^{1\over N_f}\cr
&+\sum_{l=1}^{N-2}{1\over Y_{l,N-1}}+\sum_{l=1}^{N-2}{1\over
Y_{l+1,l}}.\cr}}

\appendix{B}{Intersections}

Here we present the details of intersection number computations.
Let $X$ denote a smooth Calabi-Yau four-fold and $D$ denote
a compact smooth divisor therein.
The typical problem encountered in this paper is to compute the
intersection of $D$ with a holomorphic curve $C$ embedded in $D$.
According to
\ref\Fulton{W. Fulton, ``Introduction to Intersection Theory in
Algebraic Geometry'', Regional Conference Series In Mathematics {\bf
54}, Providence, Rhode Island (1984).},
the relevant intersection number is defined as the degree
of the pull-back line bundle $i^*{\co}(D)$ where $i:C\rightarrow D$
is the inclusion map and ${\co}(D)$ is the line bundle associated to
$D$ on $X$.
Since $X$ is Calabi-Yau, ${\co}_D(D)\simeq K_D$, therefore the
intersection
number is simply $\hbox{deg}\left(i^*K_D\right)$.
This can be applied in certain particular situations

1. $D$ is rationally ruled over a surface $B$ and $C$ is the class
of the $P^1$ fiber. The adjunction formula
\eqn\adj{
K_C=i^*K_D\otimes N_{C/D}}
shows that
\eqn\restr{
i^*K_D\simeq K_C\simeq O_C(-2)}
since the normal bundle of a fiber is trivial.
Hence $C\cdot D=-2$.

2. $\tilde D$ is the blow-up of a rationally ruled divisor $D$
along a curve $\Sigma$ in the base and $C$ is the $P^1$
fiber of the exceptional surface $H$. Note that $H$ is
a $P^1$ ruling over $\Sigma$ with normal bundle
\eqn\exnorm{
N_{\Sigma/{\tilde D}}\simeq {\co}_H(-1).}
Since
\eqn\excan{
K_{\tilde D}=K_{D}+H,}
it follows that the restriction
\eqn\exrestr{
i^*(K_{\tilde D})\simeq O_{P^1}(-1).}
Therefore, $C\cdot {\tilde D}=-1$. The generic rational fiber $F$
splits into two components $C$, $C^{\prime}$ satisfying
\eqn\fibersplit{
C+C^{\prime}=F.}
Since $F\cdot{\tilde D}=-2$, it follows that $C^{\prime}\cdot {\tilde
D}=-1$ reproducing the picture in section 1.
If $D$ is successively blown-up $n$ times along the curve $\Sigma$,
the resulting fiber has $n+1$ components $C_0\ldots C_n$ intersecting
transversely as in Fig.

\ifig\engM{The reducible fiber after $n$ successive blow-ups.}
{\epsfxsize3.0in\epsfbox{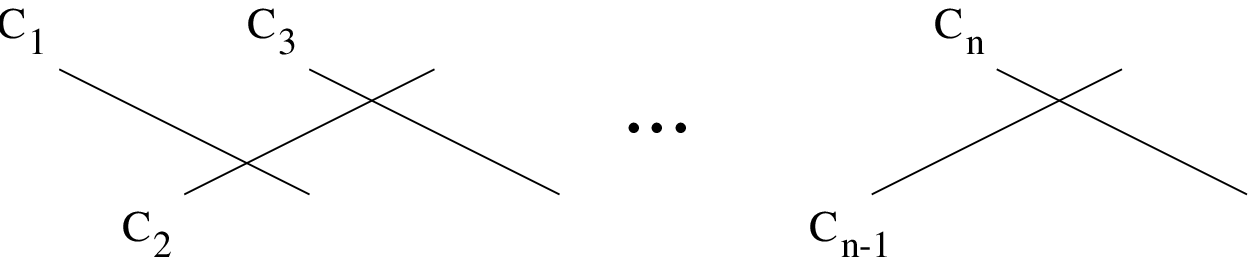}}

\noindent
A local
computation shows that the degrees of the normal bundles are
$-1,-2\ldots -2,-1$. Therefore the intersection numbers read
\eqn\multint{
C_0\cdot {\tilde D}=-1,\qquad C_1\cdot {\tilde D}=0\quad\ldots
\quad C_{n-1}\cdot {\tilde D}=0,\qquad C_{n}\cdot {\tilde D}=-1.}
This explains the correspondence between curve classes and
representation weights used throughout the paper.

\appendix{C}{$A_{N-1}+A_{N_f-1}$ collisions}

The resolution of $A_{N-1}+A_{N_f-1}$ is best understood for
Calabi-Yau threefolds \refs{\RM,\BIK,\AG,\KVb,\BJ,\A}.
The case of interest here involves a four-fold collision when
the singularities are fibered over two $P^2$ components of the
singular locus intersecting along an elliptic curve $\Sigma$.
Let $z,t$ be local coordinates such that $\Sigma$ is described as
$z=0$ in the first $P^2$ and $t=0$ in the second.
The local geometry in directions normal to $\Sigma$ is described by
the three-fold singularity
\eqn\singA{
xy=z^{N}t^{N_f}.}
Here $t$ a coordinate on the color $P^2$ which has finite volume
while $z$ is a coordinate on the flavor $P^2$ which is taken of
infinite size.

For concreteness, we consider the particular case $N=3,\ N_f=1$.
Then the singularity is
\eqn\singB{
xy=z^3t}
and the resolution involves three blow-ups.

$1.$ Blow-up the plane $\{x=z=0\}$.  The resulting space can be
covered by two affine coordinate patches $(x,y,z_1,t)$ and
$(x_1,y,z,t)$ related by $x_1z_1=1$. They map to the original $C^4$ by
\eqn\mapA{
z=xz_1,\qquad x=x_1z}
respectively. The proper transform of $X$ is characterized by the
equations
\eqn\propA{
y=x^2z_1^3t,\qquad x_1y=z^2t}
in the two coordinate open sets. The hypersurface is still singular
in the second coordinate patch, but the degree of the singularity has
decreased by one.

$2.$ Blow-up the plane $\{x_1=z=0\}$. This leads again to two coordinate
open sets $(x_1,y,z_2,t)$, $(x_2,y,z,t)$ such that $x_2z_2=1$ and
\eqn\mapB{
z=x_1z_2,\qquad x_1=x_2z.}
The blown-up space is described by
\eqn\propB{
y=x_1z_2^2t,\qquad x_2y=zt.}
This shows that we are left with an ordinary double point (conifold)
which can be resolved by a single blow-up

$3.$ Blow-up the plane $\{x_2=t=0\}$. The affine coordinate sets are
$(x_2,y,z,t)=(x_2,y,z,x_2t_3)=(x_3,y,z,t)$ with $x_3t_3=1$.
The blown-up hypersurface is smooth
\eqn\propC{
y=zt_3,\qquad yx_3=z.}

The resolved threefold can be covered by four affine open sets
$(x,z_1,t)$, $(x_1,z_2,t)$ $(x_2,z,t_3)$, $(x_3,y,t)$. The exceptional
locus consists of two rational curves $P^1(x_1)$, $P^1(x_2)$
fibered over the $t$-line yielding two rationally ruled surfaces.
Note that the step $3$ above is actually an embedded blow-up of the
second surface which resolves at the same time the ambient threefold.
Therefore, the second ruling has a double fiber localized at the point
$x_2=t=0$. Since this picture holds locally at any point on the
elliptic collision locus $\Sigma$, the global picture is the degeneration
described in section 1.

The flop of the flavor component can also be understood in this
picture. It corresponds to a different resolution in which steps
$2$ and $3$ above are replaced by

$2^{\prime}.$ Blow-up the plane $\{x_1=t=0\}$. The affine open sets
are $(x_1,y,z,t)=(x_1,y,z,x_1t_2)=(x_2t,y,z,t,)$ with $x_2t_2=1$ and
\eqn\propD{
y=z^2t_2,\qquad yx_2=z^2.}

$3.^{\prime}$. Blow-up the plane $\{x_2=z=0\}$. We have
$(x_2,y,z,t)=(x_3z,y,z,t)=(x_2,y,x_2z_3,t)$ with $x_3z_3=1$ and
\eqn\propE{
y=x_2z_3^2,\qquad yx_3=z.}

The exceptional locus consists again of two ruled surfaces
over the $t$-line with fibers $P^1(x_1)$, $P^1(x_3)$. Step
$2^{\prime}$ is an embedded blow-up of the first surface
at the point $x_1=t=0$. Therefore, the singular fiber has been
flopped from one surface to the other.

\appendix{D}{Computation of $\lambda$}

Here we compute the class $\lambda={p_1\over 2}$ for the total space of 
the canonical bundle of a rationally ruled divisor $D$ before and
after the blow-up. We consider rationally ruled divisors $D$ which can
be obtained as a projectivization of the rank two vector bundle 
$={\co}\oplus{\co}(n)$ over the base $B=P^2$. Let $X$ denote the total
space of $K_D$ and let $\pi:X\rightarrow D$ denote the natural
projection. The tangent space of $X$ fits in an exact sequence 
\eqn\exactseq{
0\rightarrow \pi^*K_D\rightarrow T_X\rightarrow \pi^*T_D\rightarrow
0.}
Therefore the total Chern class of $T_X$ is given by\foot{Since the 
degeneration can be thought as embedded in a large compact
Calabi-Yau four-fold, we do not use compact vertical cohomology on
$X$. Therefore the Thom class is simply $-s\pi^*c_1$ 
\ref\bt{R. Bott, L.W. Tu, ``Differential Forms in Algebraic
Geometry'', Springer-Verlag (1982).}.}
\eqn\totchern{
c(X)=\left(1-c_1(D)\right)\left(1+c_1(D)+c_2(D)\right)}
where $c_1(D),c_2(D)$ are Chern classes of $D$ pulled back to $X$. 
The class $\lambda$ of $X$ is then given by 
\eqn\ptr{
\lambda={1\over 2}\left(c_1(X)^2-2c_2(X)\right)=
c_1(D)^2-c_2(D).}
Therefore the computation reduces to the Chern classes of the 
divisor $D$. 
Before blow-up, the total Chern class of $D$ can be computed 
by an adjunction formula
\ref\fmw{R. Friedman, J. Morgan, E. Witten, ``Vector Bundles 
and F Theory'', hep-th/9701162.} 
\eqn\totD{
c(D)=\left(1+c_1(B)+c_2(B)\right)(1+r)(1+r+t)}
where $r=c_1({\co}(1))$ -- the ${\co}(1)$ bundle over the $P^1$-bundle
$D$ --  and $t=c_1({\co}(n))$ pulled back from $B$. 
In the present situation $H^{1,1}(D)$ is two dimensional and generated
by the divisor classes $s$ -- the section at infinity -- and $h$ -- the
pull-back of the hyperplane class of $B=P^2$. They satisfy
\ref\KM{K. Mohri,''F Theory vacua in Four Dimensions and Toric
Threefolds'', hep-th/9701147.}
\eqn\intringA{
s^3=n^2,\qquad s^2h=n,\qquad sh^2=1,\qquad h^3=0.}
Note that the zero section $r$ is $r=s-nh$. 
It is convenient to introduce the curve classes $F,H$ representing the 
generic fiber and the hyperplane class of the section $s$
respectively. 
The intersection ring is then determined by the following tables
\eqn\intringB{
\matrix{& & s & h \cr & s & nH & H \cr & h & H & F \cr}
\qquad
\matrix{& & F & H \cr & s & 1 & n \cr & h & 0 & 1 \cr}.}
Then, a direct computation shows that 
\eqn\secondch{
c_1(D)= 2s+(3-n)H,\qquad c_2(D)= 3(1-n)F+6H.}
Hence 
\eqn\lamclass{
\lambda= \left(n(n-3)+6\right)F+6H}
is even for any integer value of $n$. 

Let now $\tilde D$ denote the blow-up of $D$ along the an elliptic 
curve $\Sigma$ in the zero section. The canonical class of $\tilde D$ is 
\eqn\blowupK{
K_{\tilde D}=K_D+e}
where $e$ is the exceptional divisor. Note that $e$ is the
projectivization of the normal bundle of $\Sigma$ in $D$
$N_{\Sigma/D}\simeq {\co}(3)\oplus {\co}(-n)$. Let $E$ denote 
the curve class of the resulting rationally ruled surface. 
The intersection ring of $\tilde D$ is determined by the following 
tables
\eqn\intringC{
\matrix{&   & s & h & e \cr 
        & s & nH & H & 0 \cr
        & h & H & F & 3E \cr
        & e & 0 & 3E & 3nF-3H\cr
        &   &   &    & -3(n-3)E\cr}
\qquad
\matrix{&   & F & H & E \cr
        & s & 1 & n & 0 \cr
        & h & 0 & 1 & 0 \cr
        & e & 0 & 0 & -1 \cr}.}
The second Chern class of $\tilde D$ is of the form 
\eqn\secchblup{
c_2=aF+bH+cE}
where $a,b,c$ can be determined using 
Riemann-Roch theorem
\eqn\RRth{
\chi({\co}_W)=\int_{\tilde D} \left(1-e^{-[W]}\right)
\hbox{Td}(\tilde D)}
where $W$ is any divisor on $\tilde D$. 
This implies 
\eqn\RRthimpl{
Wc_2(\tilde D)=12\chi({\co}_W)+3W^2c_1(\tilde D)-Wc_1(\tilde
D)^2-2W^3.}
A direct computation yields
\eqn\abc{
a=3-6n,\qquad b=9,\qquad c=3n-9.}
Taking into account \blowupK\ we find 
\eqn\bluplam{
\lambda=\left(n(n+3)+6\right)F}
which is even for all integer values of $n$. 

\appendix{E}{Grothendieck-Riemann-Roch for singular varieties}

The extension of Grothendieck-Riemann-Roch theorem to singular
varieties is due to
\ref\GRRoch{J.-P. Serre, A. Grothendieck, L. Illusie,
``Th\'eorie des intersections et the\'eoreme de Riemann-Roch'',
Springer Lecture Note in Mathematics, 225 (1971); W. Fulton,
``Riemann-Roch for Singular Varieties'', in ``Algebraic Geometry,
Arcata 1974'', Amer. Math. Soc. Proc. Symp. Pure Math 29 (1975), 449;
P. Baum, W. Fulton, R. MacPherson, ``Riemann-Roch for Singular
Varieties'', Inst. Hautes \`Etudes Sci. Publ. Math. No. 45 (1975),
101.}.
Here we are concerned only with applications to singular
curves with ordinary double points, following closely
\ref\FultonB{W. Fulton, ``Intersection Theory'', Springer-Verlag,
(1984).}.
Let $C=\sum_{i=1}^NC_i$ be a reducible
curve consisting of $N$ rational components intersecting
according to either the ordinary (a) or the affine (b)
$A_N$ Dynkin diagram.
Let $\tilde C$ denote the normalization of $C$, Note that $\tilde C$
consists of $N$ disjoint rational components with a collection of
marked points that map pairwise to the double points of $C$.
According to \FultonB,\ we have
\eqn\singgenus{
\chi\left(C,{\co}_C\right)=
\chi\left({\tilde C},{\co}_{\tilde C}\right)
-\sum_{P}\delta_P}
where the sum is over all double points $P$ and
\eqn\correction{
\delta_P=\hbox{length}\left({\co}_{\tilde C}/{\co}_C\right)_P.}
For an ordinary double point, $\delta_P=1$
\ref\har{R. Hartshorne, ``Algebraic Geometry'', Springer-Verlag
(1993).}.
Since $\tilde C$ is just a collection of $N$ disjoint rational
curves we get
\eqn\genusA{
\chi\left(C,{\co}_C\right)=1}
in case (a) and
\eqn\genusB{
\chi\left(C,{\co}_C\right)=0}
in case (b).

\listrefs
\end